# Strategic Advantage in Web Tourism Promotion: An e-Commerce Strategy for Developing Countries

Arunasalam Sambhanthan (*) researches on the application of human computer interaction techniques to key application areas such as business, education and health. He holds a Bachelors degree in Technology Management and Computing from the University of Portsmouth. His honors thesis investigated on an appropriate model for achieving strategic advantage in web tourism promotion with specific focus to Sri Lankan hotels as a case study for developing countries. Presently he is a researcher at the University of Portsmouth investigating on a second life based virtual therapeutic community for patients with Borderline Personality Disorder. Sam has published his research through international conferences and peer reviewed journals. He also serves as a reviewer for the Journal of Information, Information Technology and Organizations and Electronic Commerce Research and Applications. Recently he has been invited to serve as a member of the international board of reviewers of the 2012 InSITE conference organized by the Informing Science Institute, USA.

Dr Alice Good lectures in Human Computer Interaction and Strategic Business IT, as well as previously lecturing in e-commerce. She is also supervising a number of distance learning PhDs. Research expertise in accessibility, user centered design and mental health support systems. Other research interests include e-learning and e-commerce. Currently co-coordinating a multi-disciplinary project in providing virtual and mobile support for people with mental health problems. Previous projects have looked at developing algorithms to rate the accessibility of web pages and the evaluation of learning environments. Publications include refereed book chapters, journal and conference papers. Invited reviewer for the: Interact conference in HCI, International Conference on Intelligent User Interfaces and the Transaction on Interactive Intelligent Systems Journal.

Arunasalam Sambhanthan
Department of Computing,
University of Portsmouth,
Buckingham Building, Lion Terrace,
Portsmouth PO1 3HE,
United Kingdom
arunsambhanthan@gmail.com

Alice Good
Department of Computing,
University of Portsmouth,
Buckingham Building, Lion Terrace,
Portsmouth PO1 3HE,
United Kingdom
alice.good@port.ac.uk

# Strategic Advantage in Web Tourism Promotion: An e-Commerce Strategy for Developing Countries

Arunasalam Sambhanthan, University of Portsmouth, UK
Alice Good, University of Portsmouth, UK

**ABSTRACT**

This research informs the means to develop an e-commerce strategy for web based tourism promotion of hotels located in developing countries. The study explored the aspects related to the use of information systems in web based tourism promotion, along with a focus on the organizational factors affecting the use of e-commerce strategy. Interviews were conducted with the managers of selected five sample hotels located in Sri Lanka. A structured web content analysis was undertaken for all five sample hotels to trace process level data on the e-commerce web content. Specific aspects of web content analysis include interactivity, trust, information and value adding features. Instrument for web content analysis was developed by the researcher. The outcome of research produces an outline for developing an e-commerce strategy for hotels located in developing countries to achieve strategic advantages in web based tourism promotion.

## INTRODUCTION

The unprecedented growth in Information Communication Technologies (ICTs) has revolutionized the business world. Revolutions in relation to managerial functions have resulted in an epochal impact on relevant businesses. For the survival and success of any industrial cluster, proper acquisition of technology and effective usage of its innovations have become vital elements. Tourism has been reported for its strategic utilization of ICT. Buhalis explains the impact of Information Technology (IT) in tourism industry as follows. "Information communication technologies (ICTs) have been changing the global tourism industry rapidly. The implications of the internet and other growing interactive multimedia platforms for tourism promotion are far reaching and alter the structure of the industry." (Buhalis, 2004, p. 104). Although the Information Systems / Information Technology (IS/IT) has been evolved as a strategic player and even the principle enabler in tourism promotion during the last few decades, most of the hoteliers in developing countries are still using it as a support tool for tourism promotion.

The global leading position of tourism industry has opened up a strategic window for many developing countries to boost their economy by positioning them to record global market share in tourism. In fact, the WTO predicted that in 2010, America would lose its favorable position behind Europe to Asia and the Pacific in receiving international tourists and that by 2020, Asia and the Pacific can expect around 397 million arrivals (Harris & Vogel, 2007). The growth of tourism in developing countries is growing at a faster rate than their developed counterparts. One of the primary success indicators for tourism for the developing countries is promoting their tourism attractions to the developed countries. Alongside, a survey reports 30% of the American adult population used internet to browse travel information in 2003, while the European online travels sales for the same period increased by 44% (Werthner & Ricci, 2004, pp. 101 - 105). The above study clearly shows the increasing trend of internet based tourism reservation from year 2003, over the years. Furthermore, the evolution of web platform technologies from traditional web 1.0 to interactive web 2.0 have resulted a paradigm shift in online promotion. Meanwhile Roush (2006) predicts the forthcoming invent of more advanced web 3.0 platforms. Richards (2007) indicates web 3.0 as a platform of intelligentsia, facilitating quick web searches. On the contrary, is it feasible for hotels located in the developing countries to invest a huge amount in web 3.0 technologies in the era of financial instability? Secondly, being in the cutting edge of a technological shift, will it advisable for hotels to invest on technologies going obsolete? Moreover, interactive platform is the urgent need for most of these hotels for excellence in tourism promotion. Consequently, web 2.0 with high level interactivity could be the most viable platform technology for the hotel websites. This clearly shows that the tourism businesses in the developing world experiences an emerging need for a strategic approach towards web based tourism promotion in order to enhance their web based promotion within the existing web 2.0 framework.

Tourist arrival in Sri Lanka in the last few years has been observed to be declining. Considering this, and the post-war development of the country, Sri Lankan government declared the year 2011 as National Tourism Year. It is estimated by the SLTB forecast that there will be 25 million tourists in 2016 (Antony, 2009, Para 1). Another statistic released by Sri Lanka Tourist Board (SLTB) confirms the top rank of Asia, Europe and American countries in country wise arrival rating of Sri Lankan tourism market (Annual Statistical Report, SLTB, 2008). The above findings indicate the shared chareterstics of Sri Lankan tourism industry with other developing countries, which strives to enhance the web based tourism promotion to attract tourists from the developed nations. Hence, Sri Lankan hotels could be selected as potential samples to represent the tourism businesses of the developing world.

There are two research questions presented here:

1. What is a viable strategy for developing countries to develop web based tourism promotion?
2. Looking at hotels in one area of Sri Lanka, what specific processes are required to develop a strategy?

The aim of this research is to inform an e-commerce strategy to hotels in the developing countries, which could eventually lead the hotels towards achieving strategic advantage in web tourism promotion. The research has been intended to achieve this aim through sequentially achieving the following objectives.

I. To explore the current strategies used by hotels located in Sri Lanka, to promote tourism.
II. To identify the factors affecting the usage of e-commerce in tourism promotion.
III. To inform hotels in developing countries with a process of developing an e-commerce strategy for web based tourism promotion.

**LITRETURE REVIEW**

Nowadays IS/IT is being used as a strategic player, even as the principle enabler in business success. There is also a counter argument claimed by researchers asking whether the investments on technology are sustainable. Bocij *et al* (2003, p. 514) discussed this argument and concluded that there is no significant correlation between spending on IT and profitability. However it could be counter argued that the underutilization of IS/IT may create strategic vulnerability to the organization. The European e-business market watch reports that the overall ICT utilization remains important for competitive advantage in tourism sector (E-Business watch, 2005, p 7). In contrast the success greatly depends on the approach being employed, not on the technology itself (Gretzel *et al*, 2000, p. 146). Thus developing new strategies is more productive than solely investing on emerging technologies. Apart from the above discussion, another study suggest alignment of business and IS/IT strategies as a better means to improve organizational performance (Shin, 2001, p. 227). However a thorough analysis of business needs and the existing IS/IT mix is essential to make a sense for each IS/IT investments in the current time of global credit crunch.

"Today marketers are moving towards viewing promotion/communications as managing the customer relationship over time, during the pre selling, selling, consuming and post consumption stages." (Armstrong & Kotler, 2003). Customer relationship is the process of developing and maintaining mutually beneficial long-term relationships with strategically significant markets (Buttle, 2002, pp. 2). Effective customer relationships are vital for effective communication and vice-versa. Preventing distortion in B2C communications, ensuring secure B2C communications, developing and maintaining personalized and usable communication channels, all are inclusive within the scope of CRM. Hence, a number of conceptual elements of both CRM and Promotion are not only interchangeable, but also complementary. Typically, CRM enhances promotion through adding effective C2B communication facilities to it. In contrast, the interactivity of B2C communication is specialized in promotion. Therefore, this research explores both of the above concepts, considering their complementary nature in effective online promotion. The concept of promotion is defined as follows for the scope of this research:

**Promotion is the process of initiating, nurturing and maintaining interactive communications with the market and managing the customer relationship over time, during the pre selling, selling, consuming and post consumption stages.**

The growth and diffusion of online tourism promotion has dramatically shifted the business paradigm from supply driven to demand driven. Perhaps, the use of pull promotion strategy has been subsequently increased due to the above change in business transactions. Although, it is admissible that the promotion strategy of each individual hotel is greatly depends on their unique promotional mix selection. However, maintaining and upgrading the online tourism promotion strategies is vital for any hotel regardless of the type of strategy employed.

**WEBSITE INTERACTION AND EVALUATION**

In considering the steady growth of tourism in developing countries, evaluating the functionality, usability and effectiveness of tourism based Websites is paramount. The importance of the hotel's website in its promotional value should not be underestimated, particularly given that the majority of anonymous customers initiate their communication via the website. After all, the impression building process of a particular hotel begins fundamentally at its website. Around 2001, tourism websites predominantly featured only the basic types of functionality, namely the provision of information and email reservations. However, Doolin *et al* (2002, pp. 557) claims interactivity of website as the major contributor towards the quality of service itself. Interactivity enables the effective flow of supply chain through facilitating effective demand forecast and innovative product design. In considering this, website interactivity could be marked as a 'win-win' factor in the perspective of both customer and hotelier.

Tourism marketing via the Internet has grown considerably since it commenced around 1995. Evaluation methodologies followed very soon after in 1996, where pioneer work by Murphy et al sought to evaluate early web development in tourism and hospitality sites (Murphy et Al, 1996). Since then, there has been a number of evaluation frameworks developed specifically for e-commerce websites. These include a variety of methods such as surveys, case studies, observation studies, evaluation frameworks and customer satisfaction studies. Lu et al (2002) classified ecommerce Website evaluation into four main areas: application functionality evaluation; cost benefit analysis; user satisfaction assessment and success factors identification, whereas other researchers looked at network statistics (Fletcher et al, 2002) Later research looks at recognizing the importance of correctly defining user requirements to ensure a good user experience and usability (Preece et Al, 2007). However, research into tourism Website evaluation is still limited and Law et al (2010) suggest in their paper that specific standards for tourism Website evaluation would be useful. These standards should be interdisciplinary in their approach and essentially be human centered.

Additionally, many - to - many communication facilities enhance the electronic word of mouth (e-WOM) of the hotels. Consequently, e-WOM facilitates quick tourism promotion in a low cost. Litvin *et al* (2008, p 462) indicates this as the unique contribution of e-WOM towards tourism promotion. But, the challenge relay on maintaining long term customer satisfaction, the absence of which may cause adverse effects on customer loyalty. In addition, the anonymity of e-WOM poses dependability threats to the business. Misleading messages could be easily spread out by opinion leaders using e-WOM facility. The ease of spreading messages through VTCs is another identified new threat for hoteliers. However; these facilities also could be used positively to promote the products through having an effective CRM system.

**INFORMATION QUALITY**

The actual quality of the information is an important consideration in esnuring the effectiveness of a tourism Website. Poor quality information pertaining to tourist destinations could deter potential customers, whereas accentuating the actual virtual experience can motivate consumers to visit the places. 3D virtual tours can enhance potentials'customers' impression of holiday destinations. Content needs to be rich in its presentation, timely and correct. Kozak *et al* (2005, p. 7) and Doolin *et al* (2002, p. 557) states that the web content should be regularly updated, be informative and personalized in a manner, which could directly influence the customer perceived image of destinations to create a positive virtual experience. In addition, the design of many - to - many communication facilities should be done carefully to omit customer distrust on hotel and poor customer relationship.

The importance of an effective CRM system to maintain customer trust and loyalty should also not be underestimated. It was noted that Schmidt *et al* (2008, p. 505) identified that a significant number of tourism websites for not having an effective CRM system. CRM systems can also be used to identify client profiles and thus specific strategies can be developed to promote destinations and also to present information in as specific way.

**ELECTRONIC SERVICE QUALITY (E-SQ)**

The doctoral work of Iliachenko (2006, pp 123) provides a comprehensive model to evaluate the E-SQ of tourism sites. The model of E-SQ encompasses all the critical success factors for web promotion namely, interactivity, design, information and technical features. Interactivity covers, a set of features enable the user to interact with the website. The design section covers the features make the website appealing to the customer. The information section covers company, product and tourism information while online chat, mailing list subscription, personal information storage and multimedia are categorized as value adding aspects.

This model covers all the high level elements of a web site to meet the critical mass. Moreover, the model is specifically designed for modeling tourism sites. In addition, the model effectively facilitates an in-depth analysis of the site as it provides a checklist of web quality analysis. The design section of the model is however lacking in the inclusion of a website feedback form, the number of clicks to reach main information and trust inducing design features such as, third party certificate, user guides and relevancy of domain name. The above features are vital to build an effective CRM through the website. The following section will now explore the essentialness of CRM features for web promotion.

**TRUST AND THE ROLE OF THE INTERFACE**

Corritore *et al* (2003) indicates two approaches for defining the relationship between the person who trusts and the object being trusted. More specifically, the approaches are; (1) individual to individual trust mediated though technology and (2) technology itself as the object of trust. However, web interfaces of hotels are themselves considered a 'significant object of trust', particularly in terms of the customer's perception of usability (Schlosser *et al*, 2005 & Flavian *et al*, 2005).

The application of trust concepts to online context could effectively form a new outline for trust as "The belief or willingness to believe that an online site could be relied on its goodness, strength and ability to influence customer willingness of making commitments in the form of reservation or financial transactions through it". The design features which induce belief or willingness to believe that a hotel website could be relied on its goodness, strength, ability to influence the willingness of making commitments in the form of reservation or financial transactions through it.

Online trust is a highly influential factor in ecommerce marketing. In fact the customer's initial trust plays an important role in continued usage of the site. Research shows that the use of credit cards has increased online sales to the extent that there has been a significant increase since 2005 and this attributed to the fact that customer's prefer this type of payment system (Urban et al, 2009). It is useful to explore trust inducing features of the online world in the perspective of both consumers and hoteliers.

Mc Cole (2003) points out 10 specific elements which could trigger online trust in travel sites classified under three main categories namely; ease of interaction, ease of doing business and, ease of decision making. This research is contextually relevant for the tourism domain, but lacking to include psychological design aspects such as color. A simultaneous research conducted by Head *et al* (2003) argues the lack of humanized web elements in e-commerce sites and reports a significant correlation between the humanized web design and online trust. Mere de-humanized features such as product description and tourism information will not induce trust in websites. Consumers viewed guarantees, refund, availability of product (room) and confidentiality as factors they would like the accommodation service providers included in the latter's web site (Fam *et al*, 2004). In contrast, (Stephens, 2004, p.313) reports, page layout, navigation, professional style, graphics and information content having a significant relationship with the perceived trustworthiness of the websites of small hotels while web seals do not show any correlation with perceived trustworthiness. However, the sample of the above research only includes small hotels. Furthermore, Bart *et al* (2005) reports Privacy and order fulfillment as the most influential determinants of trust for sites in which both information risk and involvement are high, such as travel sites.

The research of Yaobin *et al* (2007) on consumers' initial trust in online firms, reports that perceived usefulness, consumers' trust propensity, website security and vendor reputation have significant effect on initial trust. However, a recent study of Cheung & Law (2009) reports a significant difference between the views of online purchasers and online browsers, with regard to their perceptions of 'secure payment methods' and 'online booking and confirmation'. Furthermore, Chen (2006, pp.210) reports the following as the major factors which have statistically significant influence over the trust of consumers on an online travel site: website's reputation; characteristics; service quality; consumer's education level; consumer's overall satisfaction with similar site, and perception of risk associated with online shopping. Although the above research provides a comprehensive list, it lacks a quantitative orientation, which could be easily translated into measurable variables. Additionally, the above are more abstract factors of what influence trust on sites, yet are not a set of specifications which could guide the website design of hoteliers to induce trust on their sites. Consequently analyzing the models which facilitate the measurement of trust in an operational level would be worthwhile on designing a comprehensive research.

**MODEL OF TRUST FOR ELECTRONIC COMMERCE (MOTEC)**

The early research of Egger (2001, p 4) produced the Model of Trust for Electronic Commerce (MoTEC) identifying a number of factors such as pre-purchase knowledge, user psychology and transference medium influencing the customer perceived image by forming pre-interactional filters. MoTEC facilitates a thorough analysis of consumer paradigm before web design so that any pre-international filters could be effectively tackled. Also the model covers the entire buyer-seller interaction process and poses a strong emphasis on Customer Relationship Management (CRM). Additionally the model was thoroughly tested and refined by several researchers, so that the efficiency is high. In fact the MoTEC was built on a metaphor that people's predisposition to trust and pre-knowledge determine the initial trust. However, other factors such as consumers' comparative thinking with competitors' site, the degree of reported frauds by similar category hotels and environmental variables (Political, Economic, Socio Cultural & Technological background of the consumer) also could influence the trusting behavior. Subsequently, a criterion enabling to compare the site with other competitors is essential.

**TRUST INDUCING DESIGN FRAMEWORK – A COMPREHENSIVE MODEL**

An interdisciplinary research of (Wang & Emurian 2005) on contemporary online trust concepts incorporated a comprehensive four phase design framework to induce online trust. Graphical design aspects, overall organization and structure, Information content and social cues are the main four aspects of features. Typically, this framework addressed the weaknesses of MoTEC by adding criterions to measure company competence, interactive features and third party certifications. As a result, the model produced by (Wang & Emurian 2005) is comparatively advanced and moderated than MoTEC in research design and practice. Therefore, the trust inducing design framework of Wang & Emurian (2005) has been selected and incorporated in to the web content analysis checklist, considering its comprehensiveness and novelty.

**THE ROLE OF WEB 2.0**

The role of web 2.0 is critical for the success of any e-commerce based business initiative. More specifically Web 2.0 forms of interactive technologies; including social networking sites like Facebook have enabled ecommerce to evolve, particularly because of the power of viral marketing. Additionally customers can interact with each other and suppliers using both e-CRMs and social networking tools.

An important characteristic of Web 2.0 sites is premised on being able to incorporate various technologies and applications within the site to enhance functionality. This enhanced functionality is primarily associated with such sites being able to publish and display diverse content - content that is user-contributed, or where the site might draw information synergistically from a third party (Sellitto et al, 2010). This is particularly beneficial when it comes to web based tourism promotion. Particularly, the tourism destinations could be promoted through adding real world replications to the website. Secondly, Web 2.0 sites in allowing users to publish, display and list diverse views, opinions, pictures, sounds, and so forth, will impact content and design features that are not encountered on traditional websites (Sellitto *et al*, 2010). This is another advantage of web 2.0 in tourism promotion. Especially, electronic word of mouth, which could be better facilitated through the web 2.0. The electronic word of mouth facilitated in the form of user generated content could be an innovative way of building positive brand image of the tourism destinations.

Another study explores about group formation in social networks (Lai and Turban, 2010). The authors describe the Web 2.0 environment, its tools, applications, characteristics. It also describes various types of online groups, especially social networks, and how they operate in the Web 2.0 environment. The study reports that the social/work groups are becoming sustainable because of the incentives for participants to connect and network with other users. A discussion of group dynamics that is based on the human needs for trust, support, and sharing, regardless if the setting is a physical or virtual one, was explored. In fact, this particular aspect of social networking could well be utilized to propagate tourism related information among potential target audience. Also, this could well be utilized as a way to influence people to make purchase decisions through informal groups. In addition, a recent survey found that consumers trusted more websites with reviews than professional guides and travel agencies and far from being an irrelevance, blogs are often perceived to be more credible and trustworthy than traditional marketing communications (Akehurst, 2009).

The exploration on existing models related to this area is critical at this point. Campbell *et al* (2011) clearly shows the importance of user generated contents. Conversely, Lee *et al* (2011), reports that information diagnosticity, information trustworthiness and perceived usefulness as the critical factors for consumers' acceptance of recommendations based on EWOMS. In addition to this, Jeon *et al* (2011), propose new internet business models using Web 2.0 applications. The proposed model is classified into five basic business models namely crowd sourcing, social networking, mashup, product customization, and open market. These Web 2.0 business models can help create value by increasing efficiency in a company's value chain, delivering new customer value, and expanding the market base. This research provides a good foundation for developing new conceptual and empirical research in this area of IT research. However, the model is far advanced for developing countries, considering the under developed models of businesses located in the developing countries. Lexhaugan (2011), has identifies the dimensions and expressions of what customer perceived value is in travel and tourism websites and how it is created. The study contributes the tourism industry in a whole, but mostly focuses on the consumer side instead of the organizational side.

**INFORMATION LITERACY INITIATIVES**

Several initiatives are in place to enhance the information literacy in developing countries, by both the government as well as industry. From the perspective of government initiatives, Sri Lanka has a number of highlights which could have some contextual relevance to this research. Firstly, the Empowering 8 is an information literacy model developed by the National Institute of Library & Information Sciences (NILIS) of Sri Lanka to underpin changing education paradigms of Sri Lanka. The model focuses on promoting awareness among school children with regard to the information technologies (Wijethunga and Alahakoon, 2005). E-Sri Lanka is another initiative to implement and e-government system in the country (Davidrajuh, 2004). Nelesala is another initiative to disseminate education to rural areas through telecasters (Mozelious et al, n.d). Nenasala initiative is in place for several years and there have been considerable developments with regard to inclusive education for people with disabilities as well. Vidusuwa is a telemedicine initiative undertaken in order to increase the accessibility of healthcare facilities to the rural areas. Vidusuwa is being developed in stages with collaboration from medical professionals in the country (Chapman & Arunathilake, 2009). In contrast, there has been a considerable amount of room for development with regard to the dissemination of research literature in this regard, as most of the works done as part of the aforementioned initiatives are underreported in scholarly forums. However, it is evident that there is a remarkable base for building an information society in the country has been laid out by the aforementioned initiative undertaken in different application areas of Information communication technologies. This particular research in the context of tourism promotion will serve the purpose of laying a solid theoretical base for e-commerce based design enhancement among the businesses operating in Sri Lanka, while developing a clear policy framework for tourism industry at large across the entire developing countries.

**LITRETURE SUMMARY**

The perception of trust has been identified as a crucial factor in ensuring acceptance and usage of an ecommerce site. A perception of trust can be identified at the onset of interaction therefore a user centered approach to the design of interaction is paramount. In addition the correct acquisition and effective utilization of IS/IT is vital for the success of any business. The interchangeable and complementary nature of the conceptual elements of both promotion and CR has resulted in hotels managing both these hand in hand. Interactivity, usability and information quality of the site are vital. Where E-SQ facilitates modeling the interactivity, design, information and technical aspects, it is lacking to include trust inducing design features, website feedback form and number of clicks concept. MoTEC is somewhat lacking in testing consumers' comparative thinking and the degree of reported frauds by similar websites. However, four phased trust inducing design framework rectify the above limitations. Web 2.0 technologies have certainly impacted upon the growth of ecommerce and enabled much customer interactivity. It is essentially interactivity that is a key factor in the success of tourism web sites.

**METHODOLOGY**

The list of Sri Lankan hotels was obtained from the SLTB and western region hotels were abstracted from the list. A Google search was done for all western region hotels and the hotels holding an e-commerce site (22 sites) were short listed from the abstracted list. Five hotel samples were selected from the above two geographic zones of western region. The chosen target represents the whole of Sri Lankan hotels, but it could also be noted that the selected hotels differ with the hotels in developing countries in two different aspects. Firstly, not all the hotels located in the developing countries have large number of tourist turnover for the year. The Sri Lankan hotels do have a relatively larger tourist turnover per year. Secondly, not all the developing countries have the privilege of attracting government support to tourism industry as Sri Lankan hotels have. The Sri Lankan hotels benefits from considerable support from the government side through the tourism development authority. The post - war scenario of the country has also contributed towards the increased government concentration on the tourism industry, which was one of the mostly affected industries due the long three decades of war. However, some of the shared characteristics of these hotels with other developing countries shows similarities with other hotels located in the developing countries. For example, the investment potential of most of these sample hotels is relatively low, which resulted in a reduced investment on e-commerce based product promotion.

The methodology of this research comprises of two methods for data collection. Firstly, an interview was conducted with hotel managers. Five managers were interviewed as part of data collection. In particular, interviewing hotel managers has been selected as a viable method to elicit explorative data on how managers think about their strategies. Arguably, the quality of information gathered through interviews is richer as it allows the researcher to be interactive and responsive to the answers given by the interviewees. Also the non verbal responses and gesture of the interviewees could be observed, interpreted and analyzed using conversation analysis techniques. Consequently interviews allow the researcher to have an in-depth look of the manager's perspective. Additionally, any vague aspect of the respondents'answer could be clarified at this point. Moreover, the interviewer will have an opportunity to evaluate the effectiveness of the information, which will allow having further research within the time period. In addition the researcher can terminate and reschedule the session in the event of not gaining satisfactory responses, which is a comparative advantage of the interview method.

As indicated previously, this research is of an exploratory nature and aimed to formulate a new strategy. In fact, formulating a foundation of qualitative facts is an essential part of this research. Face to face interviews provided an opportunity to formulate the qualitative framework of this research. Also, a number of potential qualitative inputs such as the body language and facial expressions of the participants were easily traced through face to face interview. The interview was organized into two sections. Apart from this, the ability to trace in-depth qualitative data is an advantage of interview method in the context of this project. Therefore, the interview method has been chosen as the primary method to explore qualitative data.

The first section of the interview comprised of a number of open ended questions intended to trace data on how managers think about their the current web promotional strategies, existing information system strategies of the hotels to support the web promotional strategies and the factors preventing the hotels from utilizing information systems in web tourism promotion. In this section participants will be asked a set of questions and be allowed to freely express their views. Section two is organized with a number of structured questions in the form of multiple choice questions, allowing the managers to select the best suitable answer. The rationale of second section is to elicit most necessary facts on unique characteristics of hotels' information system strategy. Any unexpressed facts in the first section could be traced through these questions.

Secondly, a content analysis of the documentation took place based on the above generalization. Themes of content analysis include interactivity, information and value adding features and trust inducing design features. The documented answers were generalized based on the similarities between individual answers. More than three numbers of similar answers considered as concrete facts. However, diverse answers were considered in certain cases, where unique practices of individual hotels could not be ignored. For example, only one hotel uses CRM system with personalized services. However, this fact has not been omitted considering the absence of CRM system in other four hotels.

Instead, this unique practice has been considered as the best practice, which other hotels could contextualize and adapt. Similarly, the responses of structured questions were entered to an Excel spreadsheet application on a question basis. A generalization was made based on the highest number of responses. Aggregated responses were converted to percentage form to gain more solid quantitative facts. These facts were used to validate the interview findings. Particularly, findings on factors affecting IS/IT utilization were very useful due to the poor qualitative response to the unstructured questions.

**RESULTS**

60% of the interviewed hotels are solely using photographs to increase interactivity of their sites, while another 20% uses multimedia and the rest merely depends on emails to enhance interactivity. However, managers have shown a positive response towards creating a pleasing virtual experience. One manager expressed their special care on drafting courteous e-mails. However, emails are not adequate to create an interactive web portal as the initial impression is greatly dependent upon web interface than emails. Two managers indicated excluding irrelevant information and including standard, relevant information as their main strategy to maintain effective information portal. However, only one manager was aware of the click stream paradox and said they have designed their sites in a manner; main information could be reached within 2 clicks. The lack of concern on click rules is observed in 80% hoteliers. However, research shows tourism sites take more than three clicks to reach basic information which are not usable (Essawy 2006). On the other hand, almost all the hotels interviewed have an online reservation option in their sites, but only 40% of the hotels have an effective payment system. 80% of the hotels do not have a security risk management plan and 60% of the hotels spend below 15% of their annual income on security budget. 60% of the hoteliers have effective procedural measures to protect their customers. Particularly, providing detailed information on secure payment methods, sending confirmation emails of reservation and reducing intermediaries in payment cycle, advising customers to use only website of an accredited agent for payments are the common practices used. However, most of the managers rated decreasing security complaints, apart from two hoteliers who indicated a slow increase. Considering the lack of payment system and decreasing security complaints, the current security budget supported by the existing procedural measures is sufficient. However, considering the increasing online travel sales indicated by the hoteliers, it is wise to consider a payment system supported by a security risk management plan. The web content analysis results depicted in figure 1 shows a number of specific inadequacies with regard to the interactivity of sites.

Lack of interactive features has been identified as the main shortcoming with regard to the interactivity of western region hotels of Sri Lanka. Doolin *et al* (2002, p.560) reports, most of the New Zealand RTOs are lacking to have transactional level interactivity features such as; secure online transaction, order status tracking and integration with cooperate servers. The above futures were basically counted by the author manually. Judgment of visually pleasing interface was made based on the first impression made by the sites'interface to the researcher. The safety and security information is measured based on the amount of information presented regarding the user privacy and payment related security information.

All hotels need to adhere to unique incentive schemes to encourage online reservations. 60% of the hotels interviewed use monetary incentives to add value for their service. In fact all the hoteliers are using incentives as a value added service for their electronic service platform. However, none of the managers were concerned about using the web portal itself to add value to the electronic service. The impact of the above phenomenon could cause customer dissatisfaction in the long term. W.Kim & Kim (2004) reports information search and transaction function as the most critical aspects of a travel website. Although, a recent study argues virtual experience will have a direct effect on purchase decision (Newman 2008). On the other hand, only 2 hotels consider e-WOM as an opportunity to improve their service quality, while 2 other hotels try to hide negative feedbacks in their web portal. Furthermore, one manager indicated "we are not promising more than what we can". This is an effective strategy to reduce the unpleasant surprises during the actual visit. However, competitors' ability to commit and comply with better facilities will reduce the strategic competency of the business, although none of the hotels showed any concern towards VTC. 11 out of 80 users surveyed have recorded VTC membership. Therefore, it could be argued that the rate of VTC membership is only 13.75% of the entire sample. In contrast, the above rate was recorded from a generic user sample. Arguably, the VTC membership rate will be higher for a sample consist only tourists. In fact it is advisable to consider VTCs to reach greater amount of tourists. Litvin *et al* (2007) do however state that inters personal information sharing and e-WOM as the most important information sources, which affect the purchase decision of a consumer. In fact, an effective e-WOM strategy is a vital requirement for strategic advantage.

Figure 2 depicts the information and value adding aspects of the analyzed sites. The multimedia features were measured through the number of video clips or any related 3D virtual clips. The ability to reach a desired destination within 3 clicks has been manually checked through evaluating the number of clicks taking to reach options such as reservation, payment and room price details. In support of this, Essawy (2006, p.65) reports that customers move on to make transactions on getting in-depth information on hotel products, lack of which, will lead to non-reservation visits. The lack of tourism information in the analyzed sites therefore will most definitely have an effect on tourism promotion. Additionally, the lack of synchronized communication features will reduce the interactive communication facilities of the site and cause customer dissatisfaction.

80% of the hoteliers are using emails to provide personalized services to their customers; however one hotel chain uses a CRM technique to provide personalized service. In particular, they are maintaining an updated database shard among the entire chains. Information on the name of customer, room preferences and favorite beverages are stored in the co-operate database and the next time customer visit the hotel, he will be addressed by his first name and served with favorite food regardless of the branch of chain he visits. Another hotel collects the visiting cards of customers and sends service updates to their mail address, but this is being done manually. 60% of the hotels are using emails to retain customers for a long time. Around 80% of the hotels reply to their mails within 24 hours. One other hotel issues loyalty cards to the customers, which could be used to gain certain benefits during the next visit. However, effective CRM system is still lacking in many hotels.

Figure 3 depicts the trust inducing design features of the sites. Apart from this, the hotels records for effective security measures. However, failing to include trust inducing design features will cause negative user perception. Arguably, lack of the above trust inducing features should be the reason for user perception. In fact, the above features are essential to build trust among customers. (Wang & Emurian 2005) establishes the essence of trust inducing features in customer relationship management. However, the web site feedback form has not been observed in any websites except one.

Planning strategies involve a number of constraints. 60% of the hotels records below 25% information system integration and 15% - 25% of annual income spending on IS/IT. Although, more than 3 managers indicated global credit crunch having an impact on there IS/IT spending, 80% of the managers were unwilling to specify any factors. This might be due to their unwillingness to disclose business information. However, a number of factors specified by the managers are presented and analyzed below.

The interview results shows that only 20% of the hotels have 3-5 years strategic IS/IT planning. 60% of the managers indicated staff incompetence in IT as a main constraint affecting their IS/IT utilization in web tourism promotion. In particular, 40% of the hotels have less than 25% of their total staff exposed to IT. Only 40% of the hotels conduct periodic IT training to their staff. Another 40% of the hotels never conduct staff training programs. 60% of the hotels only spend less than 25% of their staffing cost on IT training. One manager indicated the risk of training as a deviation from the work. Additionally, a number of precedents for staff leaving the hotel after trained into IT were mentioned. The inability to cope with the increasing IT knowledge of customers with less updated technology infrastructure was indicated as another common constraint of hoteliers in IS/IT utilization. 60% of the hotels' top managements have less than average commitment towards IS/IT utilization.

**REQUIREMENT ANALYSIS**

The results show that most of the hotels do not have a web promotion strategy. Almost half of the sites researched identified as less interactive. The interactive features lacking were identified as online order tracking facility, product search, readable fonts and online payment system. Sites were identified for lacking information and value adding features. Specifically, relevant tourism information, direct link to other tourism sites, chat/customer forum, personal information storage, mailing list subscription and multimedia features. Lack of concern about VTCs was identified as a major weakness with regard to e-WOM. Proper security risk management plan and a computerized CRM system are major critical requirements for competitive focus. Apart from this, lack of trust inducing features identified as the main reason for negative user perception on security. Shortcomings were identified as a lacking of website map, website customization, user guides/manuals, statement of legal concerns, third party certificate and website feedback form. A photo enlarging option, a low and high fidelity site design and search engine optimization were identified as major accessibility requirements. Many hotels plan for a very short duration and the top management of many hotels are less concerned on IS/IT utilization. Lack of staff training, the inability to cope with the increasing customer requirements are identified as internal factors. Evolving technology has identified as a major challenge which translate the new investments as obsolete within a short period. The requirements for the next strategy have been established as the outcome of this chapter and the next chapter presents the strategy formulated for this project.

A requirement statement has been derived to precisely outline the contextual requirements of the new strategy to the reader. This statement is derived from the summary of research findings. The accumulated requirements were refined to get the final statement, which is presented below.

**An interactive, value added, informative, secure, accessible, customer focused, trust inducing website; with payment and CRM systems, which will enable the hotels to gain strategic advantage in web tourism promotion through influencing the purchase decision of customers.**

## STRATEGY

Evaluating the requirements shows there is a considerable amount of enhancements needed in the current websites along with a number of new additions. In fact, the current websites have to be redesigned in order to enhance the promotional focus in all four stages of promotion process. The main elements of the proposed strategy have been outlined in table 1 to provide a comprehensive view of the proposed strategy to the reader. A discussion is followed to justify the role of specifications, in getting strategic advantage.

As most of the hotels focus business professionals as their target groups; it is essential to be adequately effective at the e-marketplace. Busy professionals will not wish to spend much time searching travel sites or make reservations. In fact the above interactive features will support the customers to have quick searches and motivate tourists to make reservations. In addition, interactivity has been identified as a critical factor for strategic advantage in web tourism promotion (Doolin *et al* 2002). Therefore, these interactive features will increase the strategic focus.

Information and value adding features are lacking in many sites, while 74% of the users seek reliable information more than any other features. In fact, it is critical to provide relevant information and value adding features in the websites to influence the purchase decision of travelers. Therefore, including the above information and value adding features will give a strategic advantage for the hotels by increasing the annual travel income gained through web promotion. Lack of adequate information will negatively influence the purchase decision of customers (Essawy, 2006).

A significant correlation has been identified between negative user experience and users' opinion on site security. This shows that the current site design is not capable of inducing trust. In fact it is critical to include trust inducing features in the design. Additionally the above trust features have identified as the potential weakness of sites. Therefore, the above features will influence the purchase decision of browsers in a positive manner and dramatically increase the annual travel sales.

E-WOM is the next critical requirement. As discussed earlier the increasing interest on VTCs has created a new challenge for hoteliers through enabling peer to peer communication among tourists. Any negative / positive views about the service could be easily spread out by opinion leaders through VTCs. In fact hoteliers could take advantage of this facility to spread positive e-WOM about their services. Therefore advertising in VTCs and especially, placing the hotel web link in VTCs is proposed as an option with strategic focus.

A payments system is proposed however most of the hotels do not hold a security risk management plan. It is critical to have a security risk management plan to supplement the payment system. In fact, the probability of occurrences of frauds is high while introducing an e-payment system. Any potential risks related to security have to be forecasted and mitigation plan has to be developed to effectively handle security risks. Therefore, it is strategically significant to hold a security risk management plan while introducing a new payment system.

Training for staff and IT introduction programs to top management are the next critical requirements for the successful introduction and implementation of new IS/IT strategy. As identified in the research, the top management's commitment towards IS/IT is less in many hotels. However, the involvement of top management towards IS/IT have to be ensured through proper education of IS/IT to top management. Buhalis & Main (1998, pp.199) ascertain the essence of stakeholder involvement in IT adaptation of SMHOs. Additionally, the lack of staff training has been identified as the potential bottleneck in IS/IT utilization. Therefore, it is critical to have the above requirements to ensure the successful implementation of this new strategy.

Social media, online forums and portals of merged tourism information (including with booking services) are very popular these days. It would be useful if managers would consider incorporating these kinds of tools and services within their strategy.

It would be necessary to include this part in the design of e-commerce strategy in order to make it useful in wider terms. Without usage of this kind of media, promotion cannot be efficient enough in the present competitive edge. It should be noted however, usage of this kind of tools can bring strategic advantage to the users as they have an opportunity to be reach worldwide community almost without additional costs through social networking sites such as facebook, Twitter and Wyan.

## DISCUSSION

This research project report makes a contribution by providing a means to inform strategic advantage for web tourism promotion in developing countries. The outcome of this research shows the emerging need for incorporating trust, value and information aspects with interactivity. Particularly, the outcome indicates that a new model for hotel web strategy evaluation has to be developed with the focus of assessing interactivity, trust, value and information aspects. The model could be developed through a content analysis of available literature in the web evaluation frameworks. Specifically, the works of Iliachenko (2006) and Wang & Emurian (2005) have to be incorporated and re-specified for the web strategy evaluation of hotel sites.

The strengths of this research includes its research design, scope, methodology and adapting to best practices. A multi-methodological approach has been adapted to complement the weakness of each approach. Finally, the web content analysis, enabled identification of a number of process level shortcomings of the web interfaces. Particularly, each aspect of web interfaces has been thoroughly analyzed using the structured observation checklist. The results of the web content analysis were transcribed into a document and used as the building blocks of the formulated strategy. Having developed the strategy by taking in to consideration of process level data, could be featured as a contextual advantage of this research. The limitation identified relies on the interviews. Then there were two constraints faced in interviews. First was the unwillingness of managers to disclose sensitive business information. However, the conversation was manually recorded using a note book. Even though there is a chance for some facts to be missed in manual note taking, the responses were organized and converted to a word processing document within few hours of the interview took place. The risk of missing facts was reduced through this practice. Secondly, the managers were less willing to provide information on the factors affecting the use of IS/IT. However, the above risk was forecasted during the research design itself and a set of unstructured interview questions were designed to mitigate this risk. The questions were designed to cover all expected sensitive areas including the factors affecting IS/IT utilization. The third limitation of this study is its exclusion of customer service staff. No inputs were taken from hotel customer service staff. This will have a negative impact on the strategy with regard to its acceptability. Evidently, the back end service is entirely depending on the effectiveness of service provided by customer service staff. Especially, the CRM and payment systems will have both positive and negative impacts on the working pattern of back end staff. For example, the CRM system will enable them to maintain effective customer records. In contrast, the implementation of CRM system might create a fear of losing the job among customer service staff due to the automated maintenance of customer records. Customer service staff might resist the change due to the above fear of automation, which will tend them to lose their jobs. Therefore, it is essential to explore the pros and cons of new strategy, in the perspective of customer service staff, to successfully tackle the above strategic change

Jhonson & Scholes (1997) proposed a threefold model for strategic evaluation. The model comprises criteria namely, suitability, acceptability and feasibility. The suitability of proposed strategy has been proved through the data analysis. However, the strategy was not tested for its acceptability, and feasibility. Particularly, the acceptability of strategy among the stakeholders and its feasibility, in terms of financial, operational and cultural factors has not been evaluated. In addition, the financial feasibility of the strategy has to be critically evaluated in the current time of global credit crunch. Furthermore, the strategy's match with organizational culture has to be evaluated as the next critical factor. However, the above criteria cannot be evaluated based on the aggregated statistics of this research. Arguably, the financial and cultural feasibility are unique for each organizational context. Testing the acceptability and feasibility of this strategy is therefore recommended for future research. In addition, integrating the frameworks of Iliachenko (2006) and Wang & Emurian (2005) to develop a new framework for hotel website evaluation is recommended as a possible future research. A number of specific lessons were learnt in this research. In particular, selecting samples consisting of web developers. The strategy developed for this project is a preliminary step towards the increased web promotion of western region hotels of Sri Lanka, which needs to be tested and extended further. It is hoped that implementing this strategy will enable the hoteliers to increase their annual income through gaining strategic advantage.

**APPENDIX**

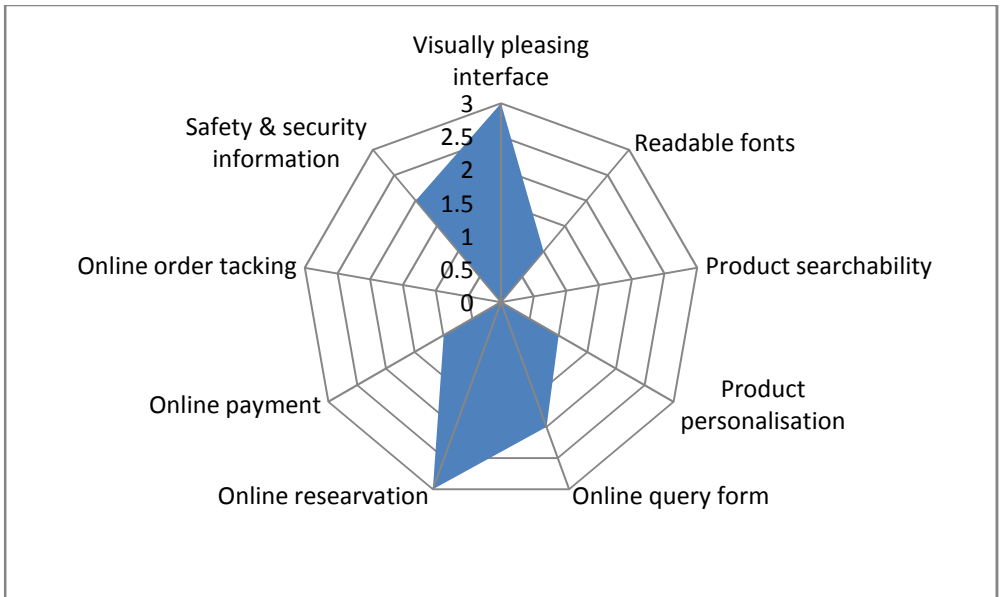

**Figure 1: Interactivity Features**

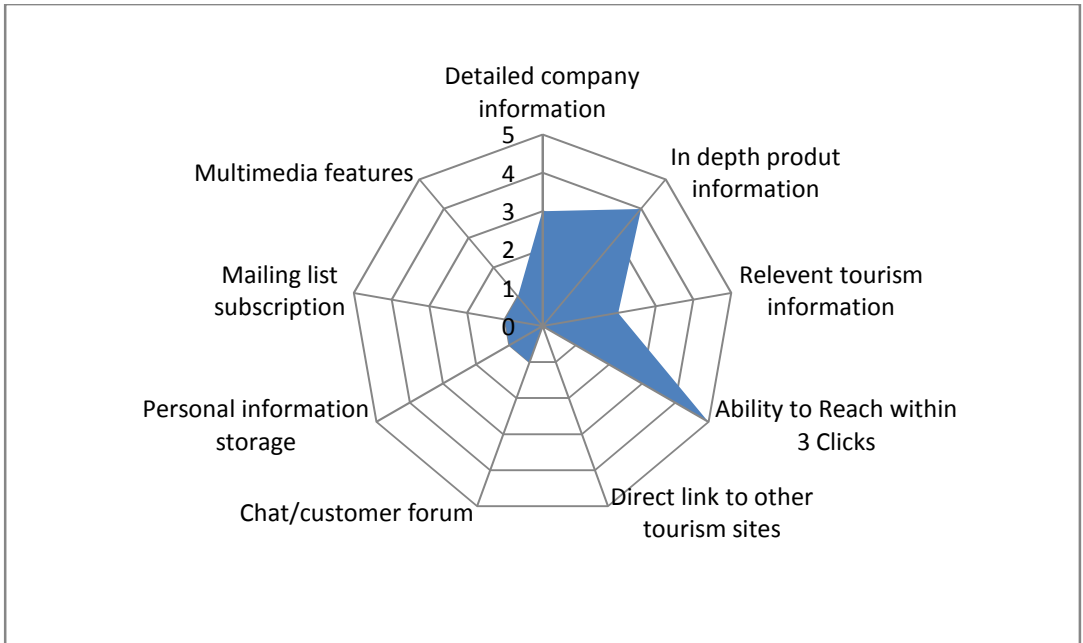

**Figure 2: Information and Value Adding Aspects**

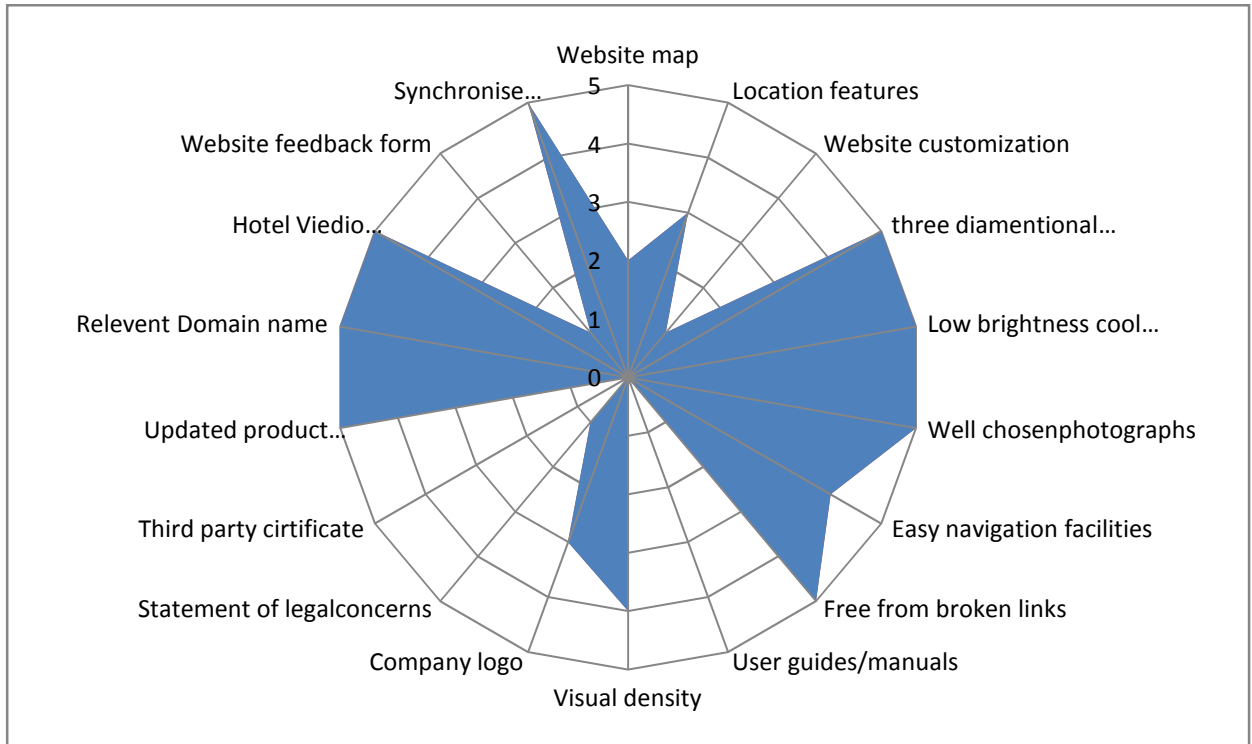

**Figure 3: Trust Inducing Design Features**

| Specification | Features | Rationale |
|---|---|---|
| Interactivity | Online order tracking | Influencing the purchase decision of browsers through enabling easy reservation and payment features |
| | Product search | |
| | Online payment system | |
| | Readable fonts | |
| Information & Vale Addition | Relevant Tourism Information | A rich site with relevant information and adding value to the service to stop customers moving to other sites |
| | Direct Link to other tourism sites | |
| | Chat/customer forum | |
| Trust Inducing & CRM Features | Website Map | Creating positive interactional filters to the site through which the trusting behavior of customers could be positively influenced |
| | Website customization | |
| | User guides/manuals | |
| | Statement of legal concerns | |
| | Third party certificate | |
| | Website feedback form | |
| | CRM System | |
| e-WOM Requirements | Advertising in VTC | Influencing Positive electronic Word of Mouth |

| Security Requirements | Security Risk Management Plan | Ensuring long term security through proper risk management |
| :---: | :---: | :---: |
| Training Needs | Staff Training on New System & IS/IT introduction sessions for top Management | Increasing promotion through enhanced staff performance with new system |
| Social Media | Social Media Marketing strategies such as Facebook, twitter, Wyan and www.bookings.com | Enhancing the electronic word of mouth of the brand image. |

**Figure 4: Proposed Strategy**